# Mixed-Dimensional van der Waals Heterostructures

*Deep Jariwala[1][†], Tobin J. Marks[1,2], and Mark C. Hersam[1,2]\**

[1]Department of Materials Science and Engineering, Northwestern University, Evanston, Illinois 60208, USA.

[2]Department of Chemistry, Northwestern University, Evanston, Illinois 60208, USA.

[†]Present address: Department of Applied Physics and Materials Science, California Institute of Technology, Pasadena, California, 91125, USA.

The isolation of a growing number of two-dimensional (2D) materials has inspired worldwide efforts to integrate distinct 2D materials into van der Waals (vdW) heterostructures. Given that any passivated, dangling bond-free surface will interact with another via vdW forces, the vdW heterostructure concept can be extended to include the integration of 2D materials with non-2D materials that adhere primarily through noncovalent interactions. In this review, we present a succinct and critical survey of emerging mixed-dimensional (2D + nD, where n is 0, 1 or 3) heterostructure devices. By comparing and contrasting with all-2D vdW heterostructures as well as with competing conventional technologies, the challenges and opportunities for mixed-dimensional vdW heterostructures are highlighted.

[\*]Correspondence should be addressed to: m-hersam@northwestern.edu

## Introduction and overview

The isolation of graphene on an insulating surface in 2004[1] gave birth to a new era of atomically thin materials in solid state electronics, which ultimately led to the search for additional two-dimensional (2D) materials.[2] The emergence of both structural and electronic variety in van der Waals (vdW)-bonded layered materials has opened new avenues for fundamental scientific studies and applied device designs. As a consequence, several combinations of distinct 2D layers have been assembled to create vdW heterostructures with varying functionalities.[3-8] These all-2D vdW heterostructures exhibit unique properties such as gate-tunability, imbuing them with potentially novel functionality versus conventional devices as summarized in several recent reviews[9-11] However, it remains a challenge to produce entire families of 2D materials and their heterostructures over large areas with high electronic quality. Furthermore, control of doping type, carrier concentration, and stoichiometry remain outstanding challenges in the majority of 2D materials, limiting the scope and progress of all-2D vdW heterostructures.[2, 11, 12]

Van der Waals interactions are not limited to interplanar interactions in layered materials. Indeed, any passivated, dangling bond-free surface interacts with another via vdW forces. Consequently, any layered 2D material can be integrated with an array of materials of different dimensionality to form mixed-dimensional vdW heterostructures. Such combinations of 2D + n-D (n = 0, 1 and 3) materials have begun to emerge and represent a much broader class of vdW heterostructures for further study. In this Review, we present a survey of mixed-dimensional vdW heterostructures, with particular emphasis on their applicability in solid-state devices. We first present a general categorization of mixed-dimensional heterostructures and introduce the classes of materials involved. We then discuss the physics of charge transport and band alignments at mixed-dimensional interfaces, while also highlighting the relevant chemistry and synthetic routes.

Finally, we discuss the applications of mixed-dimensional heterostructures in the context of solid-state devices and present a future outlook regarding their integration into mainstream technologies.

## Structure, physical properties, and fabrication of mixed-dimensional vdW heterostructures

Although mixed-dimensional vdW heterostructures have been investigated over the last two decades in organic photovoltaics , the focus here is on combinations involving at least one 2D component (Figure 1a). In earlier examples of mixed-dimensional vdW heterostructures prior to the advent of 2D materials, the interface was often extremely complex and poorly defined, making it difficult to quantify the active junction area and consequently the physical properties associated with the junction. In addition to the complex geometry, the junction interface is physically and electronically buried in most examples that do not involve 2D materials, such as organic heterojunctions[13] and composites composed of nanowires or nanotubes[14] with 3D or 0D materials. In contrast, a continuous, atomically thin sheet of material presents a more well-defined interface geometry that is amenable to quantitative characterization. Furthermore, the low density of electronic states and lower values of relative permittivity in 2D materials (as compared to bulk) renders them semi-transparent to applied electric fields normal to the 2D plane. Consequently, the junction interface can be modulated as a function of an applied gate voltage, leading to fundamentally new device phenomena in addition to providing an additional knob for tuning device properties.

Hence, we discuss here three distinct mixed-dimensional heterostructure geometries, namely 2D-0D, 2D-1D, and 2D-3D (Figures 1 b-d). The 2D components such as graphene, hexagonal boron nitride (h-BN), and transition metal dichalcogenides (TMDCs) (Figure 1a) have been thoroughly investigated.[2, 9, 10, 12, 15, 16] The 0D materials class primarily consists of fullerenes,[17] small organic molecules,[18] and quantum dots,[19] while the 1D materials class includes semiconducting single-walled carbon nanotubes (SWCNTs),[20] semiconducting organic polymers,[18] and inorganic semiconductor nanowires.[21] Finally, our discussion of the 3D materials class will focus on crystalline bulk Si, Ge, III-V, and II-VI semiconductors as well as semiconducting amorphous oxides (such as indium-gallium-zinc oxide (IGZO)[22] and amorphous Si). Figure 2 depicts examples of each materials class.

**Physical Properties** In contrast to conventional, epitaxially grown heterostructures, the interface in a mixed-dimensional vdW heterostructure is relatively more complex and less constrained, due to the absence of the need for lattice matching. Furthermore, in addition to the usual discontinuity in band structure and resulting potential energy barriers, the density of states (DOS) also undergoes an abrupt transition, which has several observable implications, such as additional resistance at the junction that results from the change in the number of conductance channels.

Typically, point and line defects in a covalently or ionically bonded interface are the source of trap states, that have energy levels within the gap of the constituent semiconductors, giving rise to Fermi level pinning and non-radiative carrier recombination. While most vdW interfaces (including many of the all-2D heterostructures) are immune to this class of defects, they are still affected by trap states that originate in substitutional defects and charge impurities within the atomic layer.[23] Similarly, disorder, contamination, and charged impurities in the overlying or underlying layers can induce interface states, unless they are effectively screened.[23]

At strongly bonded interfaces, sufficient hybridization of atomic orbitals exists to permit carrier delocalization; this hybridization enables charge transfer across the interface, resulting in band bending and formation of a built-in potential. In contrast, for materials interacting solely via vdW forces, the atomic and molecular orbitals on the surface and edges are completely saturated and thus, the orbitals at the interface do not undergo hybridization. Therefore, the out-of-plane carrier mobility is typically lower than the in-plane one, which minimizes delocalization of carriers and diffusive transport across the vdW interfaces.[24, 25] Charge transfer, however, still occurs via tunneling or hopping. The net effect is that conventional interfacial concepts (such as depletion regions), which are based upon carrier diffusion models,[26] need to be reconsidered for vdW heterostructures. Furthermore, in vdW-bonded materials (such as organic molecules and polymers), charge carriers are often described as a combination of electronic charge coupled with a nuclear distortion (e.g., polarons). Therefore, at organic-2D interfaces, the interfacial transport often involves hopping of charges from polaron states in organic materials to free carrier states in crystalline 2D materials.[27, 28] Mixed dimensionality further contributes to interfacial disorder, particularly in the 2D-0D and 2D-1D cases, since 1D and 0D materials commonly exist in film form[20, 29] with imperfect ordering, resulting in an ensemble of nanoscale heterojunctions as opposed to an electronically homogenous 2D interface.

**Fabrication** Assembly of disparate materials with precise control over interfacial properties is critical to the operation and performance of solid-state devices. Mixed-dimensional vdW heterostructures present unique challenges and opportunities in this regard. Unlike bulk inorganic or all-2D vdW heterostructures (where all components can be grown in the same reactor by switching precursors in a single growth process), the synthesis of two materials with disparate atomic arrangements and ranges of thermal stability in a single process is exceedingly challenging. Therefore, most mixed-dimensional vdW heterostructure examples involve the deposition or synthesis of the more thermally and chemically stable component, followed by deposition of the less stable component. For the case of 2D/organic semiconductor heterojunctions, the 2D material is invariably synthesized first, followed by thermal evaporation of small organic molecules[30] or spin/dip-coating of polymeric materials from solution.[28] Solution-phase layer-by-layer assembly or wet transfer techniques have also been used for 0D quantum dots[31, 32] and 1D carbon nanotubes.[33, 34] Conversely, 2D-3D heterostructures typically require growth or transfer (achieved by either dry or wet transfer using a polymer support, which is then dissolved) of the 2D material on top of the bulk 3D semiconductor. Direct chemical vapor deposition (CVD) of 2D materials on a 3D semiconductor (e.g., graphene on Ge[35] and $MoS_2$ on GaN[36]) as well as the converse (direct synthesis of gallium nitride (GaN) on graphene[37] and h-BN[38]) have been recently reported. Finally, 2D polymeric materials have been self-assembled or polymerized using small organic molecule precursors.[39-41] Layered crystals that are both chemically and mechanically stable can be synthesized by this route directly on a 0D, 1D 2D or 3D solid[39, 42] and even liquid interfaces[40, 43]. They can also be mechanically exfoliated into individual sheets onto 3D surfaces[39, 41] in contrast to the solution assembly techniques discussed above.

With the synthetic and assembly advances described above, mixed-dimensional vdW heterostructures have been successfully demonstrated in multiple classes of solid-state devices including logic devices, photodetectors, photovoltaics, and light-emitters. In each case, the integration of a 2D material with a non-2D material has resulted in significant advances, either in performance or function. The remainder of this article focuses on critically reviewing these developments by device class. Current best practices, advantages, and limitations will be highlighted in each case; relevant comparisons will also be made with corresponding all-2D vdW heterostructures as well as competing state-of-the-art technologies.

## Field-effect and logic devices

Logic devices are the fundamental building blocks of modern microelectronic circuitry. Fundamentally, a logic device is a switch that exhibits at least two distinct and measurable states.

Electronic logic devices are the most prevalent, with the field-effect transistor (FET) being the most common foundational circuit element.[26] Low-dimensional materials hold promise for shrinking device dimensions to the molecular scale.[29] For example, significant effort has been devoted to organic[18] and molecular[44] electronics, as well as the (more recent) integration of 2D materials with organic semiconductors; although promising, these approaches have not managed yet to compete favorably with conventional silicon electronics.

The basal planes of 2D atomic crystals provide atomically flat and inert surfaces, ideal for ordered self-assembly of organic small molecules.[30] In particular, the isolation of atomically thin graphene and h-BN on insulating substrates has permitted their integration with organic small molecules, in order to create functional devices. Using 2D h-BN as a substrate (Figure 3a), crystalline submicron thick films of rubrene have been templated, thereby enabling the fabrication of high-performance organic FETs (OFETs) with carrier mobilities exceeding 10 $cm^2/V.s$.[45-47] Prior to this templated growth, this level of performance in organic small molecules was only possible in large (mm size) single-crystal samples. Graphene has played a similar role for electrical contacts, wherein the highly delocalized graphene electronic structure leads to preferential face-on stacking for most $\pi$-conjugated organic small molecules (Figure 3b).[48, 49] Both graphene[50-52] and graphene-coated metals[53] have been used in pentacene FETs, each resulting in mobility increases ranging from 10-fold to 100-fold due to improved carrier injection. Solution-phase ordered assembly of small organic molecules on graphene has also been achieved[52, 54, 55] and has been successfully used for high-yield liquid-phase exfoliation of layered crystals.[56, 57] Furthermore, solution-based blends of semiconducting polymers and small molecules with graphene[58] and $MoS_2$[59] have achieved higher crystalline order in the channel, affording mobility increases ranging from 5-fold to 100-fold.

In addition to enhancing OFET performance, templated organic semiconductor films atop 2D crystals facilitate additional field-effect device designs. For instance, since graphene incompletely screens vertically applied gate electric fields, a vertical stack of graphene/organic/metal on a back-gated insulator can be operated as an FET using out-of-plane carrier transport. This vertical FET (v-FET) geometry solves a long standing problem of scaling down OFETs to submicron channel lengths, ultimately leading to OFETs with simultaneously high on-current densities and high on/off current ratios (Figure 3c).[60] In a vertical OFET, the gate electric field allows modulation of the Fermi level in both the graphene electrode as well as the organic semiconductor, resulting in simultaneous increase (decrease) of barrier height and increase (decrease) in barrier width in the off (on) states. Vertical OFETs have been demonstrated with p-type small organic molecules,[60] n-type fullerenes,[61] and conjugated polymers,[62] thus enabling complementary electronic circuits.[62, 63]

The v-FET concept originates in all-2D graphene/TMDC/metal (graphene) heterostructures[5, 7] and was subsequently extended to other passivated semiconductor surfaces, including organics and amorphous oxide semiconductors (AOSs).[64] Since AOS thin films can be deposited over arbitrary substrates via solution-based or sputtering methods, large arrays of a-IGZO v-FETs have been demonstrated with CVD graphene electrodes.[65, 66] However, unlike organic semiconductors, AOSs have much higher relative permittivity and thus the current modulation is achieved solely from tuning the graphene Fermi level and consequently the Schottky barrier height.

The tunability of the Schottky barrier height with a graphene contact was the precursor to the above discussion on v-FETs. This concept was first demonstrated in junctions of graphene with hydrogen-passivated Si giving rise to the concept of field-effect (i.e., gate-tunable) Schottky barriers or barristors.[67] Most prominently, a nearly ideal Schottky diode behavior was observed when CVD graphene was integrated with hydrogen-passivated p-type Si. In this case, the graphene/p-Si Schottky barrier height is gate-tunable, yielding current on/off ratios of ~$10^5$ under forward bias, which is suitable for digital logic applications (Figure 4a).[67] The barristor concept has subsequently been extended to organic semiconductors such as pentacene.[68]

The above discussion has been limited to graphene as a gate-tunable 2D contact in Schottky barrier or v-FET devices. However, mixed-dimensional p-n heterojunctions have also been recently fabricated and employed for logic applications. In this case, graphene is replaced with a TMDC monolayer, as in the early example of a gate-tunable p-n heterojunction based on p-type semiconducting SWCNTs and n-type single-layer $MoS_2$.[34] This device shows tunability in the rectification ratio by over 5 orders of magnitude in addition to exhibiting a unique anti-ambipolar transfer response (Figure 4b). The change in the polarity of the transconductance in the anti-ambipolar transfer response enables applications in analog circuits such as frequency multipliers and keying circuits that are widely used in modern wireless communication technology.[33, 34] The anti-ambipolar response has been further generalized to other semiconductor systems such as AOSs[33] and organic small molecules.[69, 70]

A single global gate in these p-n heterojunctions induces the same carrier type in all of the vertically stacked layers. Hence, it is difficult to form high doping levels on both sides of the p-n heterojunction unless one component is heavily doped in equilibrium. Toward this end, the doping level in 3D semiconductors can be precisely controlled by substitutional dopants. Therefore, gate-tunable 2D-3D p-n heterojunctions[71, 72] present a unique opportunity to realize highly doped p-n heterojunctions and resonant tunneling devices for logic applications. A noteworthy recent example is a p-n heterojunction based on heavily doped p-Ge and bilayer n-$MoS_2$ (Figure 4c). Upon gating the $MoS_2$ into the highly n-type doping regime, direct tunneling occurs from the Ge valence band to the $MoS_2$ conduction band (Figure 4d), resulting in a subthreshold swing below 60 mV/dec at room temperature.

Modulation of the insulating tunnel barrier height by biasing the sandwiched 2D material also allows exponential control over tunneling currents. This principle has been applied to hot electron transistors (HETs) based on 2D/3D heterojunctions[73] to achieve high and low conductance states. HETs are three-terminal (i.e., emitter, base, and collector) heterostructure devices where the 2D material (base) is sandwiched between ultrathin insulating tunnel barriers (Figure 5a). The emitter is typically a doped 2D semiconductor, while the collector is a 3D metal. A voltage applied to the base allows control over electrons tunneling from the emitter to the collector (Figure 5b). HETs employing a graphene base and heavily doped Si were the first to be proposed[74] and then demonstrated.[75] While early designs lacked a high on-state conductance and current gain (i.e., collector to emitter current ratio), the issue of low-gain was recently resolved using CVD-grown $MoS_2$ as the base material.[76] However, low on-state currents still prevent high frequency operation, although using bilayer oxide tunnel barriers leads to significant improvements.[77] The structure and operating principle of HETs is similar to that of heterojunction bipolar transistors (HBTs) with early HETs comprised of insulators sandwiched between metals.[78] Since both conventional HETs and HBTs possess cut-off frequencies limited by the base transit time, atomically thin base materials hold significant promise for the future. With simulated cutoff frequencies in the few THz regime,[73] effectively integrating 2D materials into these device concepts may enable them to be competitive with high electron mobility transistors (HEMTs) that are currently widespread in wireless communication technologies.

Tunnel barriers are similarly important for logic devices using alternative state variables, such as spin. Spin injection in semiconductors has been conventionally achieved using tunnel barriers at the ferromagnetic metal contacts to accommodate the large difference in conductivity between the metal and semiconductor. For example, $AlO_x$ and MgO have been previously employed as tunnel barriers but suffered from trapped charges and interlayer diffusion issues. In contrast, graphene offers ideal impermeable barriers between ferromagnetic metals and bulk semiconductors.[79] The relatively poor graphene out-of-plane versus in-plane conductivity creates a smooth transition between the ferromagnetic metal and semiconductor, yielding a resistance-area product that is three orders of magnitude lower than oxide tunnel barriers. The net result is room temperature spin injection and extraction in NiFe/graphene/Si heterojunctions (Figure 5c). Further exploiting this phenomenon in Si nanowires results in clean magnetic switching characteristics in spin valve devices.[80]

## Light sensing, harvesting, and emitting devices

The above section primarily focused on electronic transport phenomena and applications of mixed-dimensional vdW heterostructures. However, since mixed-dimensional vdW heterostructures tend to have higher optical absorption cross-sections than all-2D vdW heterostructures, applications in optoelectronics and photon-harvesting technologies are also promising. In this regard, recent examples of mixed-dimensional vdW heterostructures in photodetector, photovoltaic, and light-emitting applications are reviewed below.

**Photodetectors** Incident photons with energies exceeding the semiconducting band gap create bound electron-hole pairs (i.e., excitons) or free carriers depending on the exciton binding energy. Separated free carriers that are collected at the electrodes lead to detectable photocurrents. Two major categories of semiconductor-based photodetectors are phototransistors and photodiodes.[26] Phototransistors are field-effect transistors with a photoactive channel that rely upon the increase or decrease in channel conductivity upon illumination. Their sensitivity can be tuned by orders of magnitude by depleting or accumulating the channel using a gate electrode, while their response time is limited by the carrier mobility and channel dimensions. Likewise, the responsivity (electrons out per photon in, in units of $AW^{-1}$) of a phototransistor can be amplified upon application of a drain bias.

In atomically thin semiconductors, the total photocarrier generation efficacy is limited by the level of optical absorption. Conventional graphene phototransistors have limited responsivity (~$10^{-2}$ $AW^{-1}$)[81] due to weak optical absorption. However, heterojunction phototransistors where the graphene FET is sensitized with another semiconductor, can overcome this limitation. This approach was first adopted using semiconducting quantum dots (QDs) as sensitizers in graphene FETs[32]; QD optical absorption leads to photoexcited charge carriers with the holes transferred into the graphene (Figure 6a) and then efficiently collected from the graphene FET channel due to high carrier mobilities, ultimately affording responsivities exceeding $10^7$ $AW^{-1}$. This responsivity can be tuned by seven orders of magnitude as a function of the applied gate voltage, while the spectral response is controllable by varying the QD size[32, 82] or by employing another sensitizer such as small organic molecules.[83] The same approach in an all-2D heterostructure uses 2D $MoS_2$ as the sensitizer. However, the total optical absorption and spectral selectivity are limited by the ultrathin nature and availability of 2D sensitizers.[84] Phototransistors based on TMDCs as the base channel material have also been demonstrated, where the primary benefit is the exponential tunability of the channel conductance and, consequently, of the photoresponse. The use of organic small molecule sensitizers, can increase the limited optical absorption, yielding significant enhancements in responsivity even for monolayer TMDC channels.[85, 86]

Phototransistors based on the heterostructure cases discussed above, are inherently slow in temporal response ($10^1$ to $10^{-6}$ sec) due to limitations in FET channel mobility or the rate of charge transfer from the sensitizer to the channel. Photodiodes can overcome this issue since the excited photocarriers are accelerated out of the junction depletion region by the built-in electric field, resulting in high frequency operation (up to ~100 GHz, translating to a response time of ~$10^{-11}$ sec) suitable for optical communications. In contrast, the maximum photodiode responsivity is typically limited to 1 AW$^{-1}$. Photodiodes are typically operated in the reverse bias mode as it allows efficient extraction of the photoexcited carriers due to the larger electric field at the junction. In some cases, however, where the dark (leakage) current under reverse bias is high, photodiodes can be operated under zero bias (photovoltaic) mode to maximize signal to noise ratio.

Schottky diodes between graphene and crystalline bulk semiconductors were the first to be investigated for high-speed photodetection; early demonstrations of graphene-Si Schottky diodes exhibited response times of 3-5 ms,[87] which was subsequently decreased to ~ 23 µs in graphene-Ge diodes.[88] The ideality factors, response time, and detectivity of Schottky photodiodes strongly depend on the interface and graphene quality,[89-91] while the spectral selectivity of graphene-semiconductor Schottky diode photodetectors is determined by the semiconductor band gap. Heterojunction photodiodes have been demonstrated with a broad range of spectral selectivity ranging from ultraviolet using ZnO nanowires[92] to visible frequencies with cadmium chalcogenide nanowires/nanobelts[93] and Si quantum dots[94] to near-infrared using InAs nanowires.[95] Photodiodes based on p-n heterojunctions have likewise been developed and evaluated. For example, atomically thin TMDCs have been employed as the n-type layer with a-Si,[96] organic small molecules,[97] and SWCNT films[34] as the p-type layers. SWCNT-MoS$_2$ p-n heterojunction photodetectors show concurrent spectral response in the visible and near-infrared with an instrumentally-limited response time of 15 µs[34] (Figure 6b).

Electronically coupled mixed-dimensional vdW heterojunctions often facilitate greater optical absorption or photoexcited carrier separation than their all-2D counterparts. To achieve the former, light can be directed in-plane to the 2D material using optical waveguides[98] or Fabry-Perot microcavities[99] where the interaction is enhanced by multiple light passes through the 2D layer. Specifically, Si waveguides coupled to graphene phototransistors have been used to achieve high-frequency broadband optical modulators[100] and high-speed photodetectors.[101, 102] Optical modulators have similarly been demonstrated with modulation frequencies exceeding 1 GHz for a broadband spectrum of 1.35-1.6 μm.[100] Likewise, graphene phototransistors coupled with Si waveguides have yielded broadband responsivities in excess of 10 mAW$^{-1}$ at zero bias (Figure 6d) and cut-off frequencies > 10 GHz (Figure 6d inset).[102] The cut-off frequency, generally limited by mobility and parasitic capacitance, can be further enhanced to 40 GHz using graphene encapsulated with h-BN.[103] Si waveguides have also been employed in graphene-Si Schottky junction photodiodes to increase the responsivity.[104]

**Photovoltaics** Among mixed-dimensional vdW heterostructures, graphene-Si heterojunctions have been the most heavily investigated for photovoltaics thus far. The earliest devices were fabricated by etching a window in thermally grown $SiO_2$ on Si, followed by graphene transfer (schematic in Figure 7a), with power conversion efficiencies (PCEs) of ~2 %.[105] Later generation devices involved careful silicon passivation,[106] antireflective coatings,[107] and chemical doping of graphene[108] to maximize the work function difference and carrier collection efficiencies, which increased the PCE to ~8%. Since the photovoltaic performance of these planar 2D-3D heterojunction devices is limited by the graphene-Si contact area, Si micropillar arrays have been integrated with graphene to improve carrier collection (Figure 7b) and PCE.[109] Maximizing PCE above 10% requires concurrent optimization of several parameters including doping profiles,[110, [111] electron and hole transport layers,[110, 112] light trapping strategies,[109, 111] graphene layer number,[109] and surface passivation.[110, 113] The atomically thin nature of graphene also allows tuning of the doping profile with a gate electrode. Specifically, gate-tunable Schottky junction solar cells have been demonstrated in GaAs[114] and zinc phosphide[115] heterojunctions with graphene, with the highest PCEs reported exceeding 18%.[114] A variety of other low-dimensional inorganic semiconductors ranging from 1D CdSe nanobelts[116, 117] to 0D CdTe QDs[118] have also been integrated with graphene, although the PCEs for these systems remain at or below 3%. Several other studies have explored Schottky junction solar cells based on graphene and other semiconductors as discussed in a recent review.[119]

In addition to Schottky junctions, mixed-dimensional vdW heterostructures have also been incorporated into p-n junction solar cells. Early reports combined micromechanically exfoliated n-type $MoS_2$ flakes with passivated p-type Si,[120] whereas later designs employed large-area CVD-grown $MoS_2$, yielding PCEs of ~5 % with open circuit voltages of ~0.4 V (Figure 7c).[121] The performance was further increased using direct band gap GaAs and monolayer h-BN as an interfacial layer. In particular, an optimized PCE exceeding 9% was achieved in a GaAs/h-BN/$MoS_2$ device, where the $MoS_2$ was depleted with an electrolytic gate.[122] Since $MoS_2$ is typically n-type, integrating it with p-type small molecules also results in p-n heterojunction diodes with photovoltaic characteristics.[69, 70] A recent demonstration of a photovoltaic effect in a p-pentacene/n-$MoS_2$ device in a lateral, gate-tunable geometry (Figure 7d) revealed the potential for open circuit voltages great than 0.5 V[69]; this work also suggests the potential of $MoS_2$ as an alternative acceptor in bulk heterojunction photovoltaics. Integrating photonic architectures[123] and waveguides[124] for light-trapping also hold promise for enhancing optical absorption in a layered heterojunction geometry.

**Light-emitting devices** When a photovoltaic cell is operated under forward bias, electrically injected electrons and holes can radiatively recombine in the semiconductor to emit light. As the emission intensity is typically orders of magnitude higher in the case of direct band gap semiconductors compared to indirect band gap ones,[26] monolayer TMDCs that fall in this category are suitable for light-emitting applications,[125, 126] with several recent reports, demonstrating light emission from all-2D TMDC heterostructures.[127, 128] Here, we discuss efforts of integrating 2D materials with bulk 3D semiconductors in light-emitting devices. Early efforts using p-Si/n-MoS$_2$ heterojunctions as discussed above,[120] when operated under forward bias, resulted in light-emission due to injected holes recombining with electrons in MoS$_2$ (Figure 8a). The electroluminescence spectrum closely resembles that of photoluminescence, suggesting that the emitted light is due to exciton recombination in MoS$_2$.[129] Besides p-Si, efforts have also been made to integrate MoS$_2$ with p-GaN.[130] Since GaN is a wide band gap blue-emitting material, its integration with the comparatively smaller band gap, red-emitting MoS$_2$ results in a broadband emission LED; using an insulating alumina barrier between the GaN and MoS$_2$ minimizes carrier diffusion and increases emission efficiency.[130]

Integrating 2D materials in non-emitting LED elements also results in significant performance enhancement. For example, graphene and h-BN have been employed as buffer and seeding layers, respectively, for optimized growth of the emitting-layer in GaN LEDs.[37, 38] The resulting high-quality GaN films serve as the substrate for the growth and fabrication of indium gallium nitride (InGaN) multiple quantum well (MQW) LEDs (Figure 8b). Unlike the conventional sapphire substrates for epitaxial GaN growth, ultrathin graphene and h-BN interlayers allow subsequent transfer of the fully fabricated LEDs onto arbitrary substrates (Figure 8b-c).[37, 38] The same approach has been adopted to grow vertically aligned arrays of core-shell ZnO-GaN nanowires, which are then used as substrates for InGaN MQW nanowire LEDs on flexible substrates.[131] Graphene also helps promote carrier injection in LEDs; specifically, the interface of a 1D nanowire with 2D graphene provides a hot spot for current injection and radiative recombination as demonstrated in control nanowire/graphene heterojunction devices.[132] This concept has also been extended to organic LEDs (OLEDs). Of particular note is the use of highly doped, low sheet resistance CVD graphene as the anode in flexible white OLEDs,[133] which outperform indium tin oxide (ITO) anodes in current efficiency. A graphene interlayer between a GaN LED and the sapphire substrate also helps reduce the thermal boundary resistance, providing improved heat dissipation (Figure 8d).[134]

Efforts to design light-emitting devices with semiconducting TMDCs as the principal light-emitting element, have been confounded by the low quantum efficiency in these 2D materials ($\sim 10^{-5}$), which has been attributed to surface states and point defects. Recent work has shown however, that surface passivation methods can dramatically improve the quantum efficiency.[135] In parallel, significant effort is currently being expended to improve their crystal quality, identify alternative 2D materials with intrinsically higher quantum efficiencies (e.g., layered perovskites[136]), and develop innovative device designs that maximize radiative recombination in monolayer materials.

## Summary and outlook

From the early studies of 2D materials for electronic and optoelectronic applications, it has been clear that integration with other materials and controlling the resulting interfaces would be a major challenge for technologically relevant applications. In particular, currently demonstrated all-2D vdW heterostructure devices suffer from performance limitations and/or cost disadvantages, compared to pre-existing conventional approaches. Therefore, in the near term, complementing existing semiconductor technologies (such as bulk Si and III-V materials, organics, nanowires/nanotubes, quantum dots, and amorphous oxides) with 2D materials appears to be the most promising strategy. In this regard, mixed-dimensional vdW heterojunctions have shown significant early promise, although several questions remain unresolved.

For logic devices, while much effort has been devoted to proof-of-concept mixed-dimensional vdW heterostructure devices at low frequencies, there is limited compelling experimental data on the high-frequency performance of vertical heterojunctions. Given their ultrashort channel lengths, cutoff frequencies are predicted to be in the THz range; issues related to material quality, leakage currents, and device layout have prevented the full realization of this high-speed potential. For similar reasons, efforts to achieve higher levels of vertical integration have also encountered limited success. Improvements in assembly and interfacial control are thus likely to enable significantly more complex gate-tunable electronic systems that involve multi-layer and multi-terminal vdW heterostructure devices.

For mixed-dimensional vdW heterostructure photodetectors and photovoltaics, significant progress has been made in integrating graphene with Si and other low-dimensional semiconductors. For photodetectors and photovoltaics that employ 2D TMDCs as the optoelectronically active element, the combination of nanophotonic and light-trapping elements with mixed-dimensional vdW p-n heterojunctions will contribute in enhancing the optical absorption of the atomically thin films. This approach will not only maximize the generation of photoexcited carriers but will also allow efficient and fast collection, thus improving the response time for photodetectors and the power conversion efficiency for photovoltaics. However, there are underlying tradeoffs between maximizing absorption and minimizing thickness for efficient carrier collection as well as minimizing direct tunneling between cathode and anode. Particular attention must therefore be given to the carrier collection geometry; while the light trapping design must be insensitive to incident angles, the carrier collection for ultrathin heterostructures must be in-plane to maximize shunt resistances and enhance open circuit voltages. Conversely, for thicker heterostructures, out-of-plane carrier collection is more efficient without compromising shunt resistance and open circuit voltage. Typical light trapping designs in thin films involve the cladding with high index media to confine more modes in the active region.[137] Combining optical elements of large band-gap, high-index 3D semiconductors (such as GaN, AlGaN and ZnO) with 2D TMDCs would present an effective approach in such designs. Likewise, heterostructuring with photonic crystals and dielectric Bragg mirrors made from large band-gap, high-index semiconductors are also likely to prove effective at light trapping in 2D semiconductors with band-gaps in the visible portion of the electromagnetic spectrum.

Similarly, efforts towards efficient light-emitting mixed-dimensional vdW heterostructure devices will require innovative device designs to maximize radiative recombination within the monolayer semiconductor. Towards this end, advances in surface passivation and encapsulation to minimize non-radiative recombination are the necessary first steps.[135] Unlike photovoltaic cells and photodetectors, light-emitting devices will require active layer architectures that maximize the interaction of minority carriers. In monolayer thick heterointerfaces, this is a particularly challenging problem when the carrier transport is out-of-plane since a large proportion of minority carriers will pass through the semiconductors and get collected at the metal electrode. A promising strategy is to have a 1D lattice of repeating heterojunction stacks akin to MQW structures.[128] Injecting one carrier type out-of-plane and the other carrier type in-plane will also increase minority carrier interaction in the active layers. Coupling such architectures to optical cavities will enable more advanced light sources such as electrically pumped lasers. Recent developments on optically pumped control of spontaneous and stimulated emission from monolayer TMDCs in photonic crystal cavity and microdisk resonator heterostructures[138, 139] are promising steps in this direction.

Large-area integration of 2D materials with materials of other dimensionality is likely to have a sizable impact on semiconductor technologies. While reliable heterostructures have been demonstrated from micron-scale mechanically exfoliated samples, efforts to reproducibly synthesize 2D materials at the wafer-scale are still in their infancy. Furthermore, as opposed to conventional semiconductor technology, most device types discussed above have < 5 nm interface thickness, which implies that single crystalline film quality and thickness uniformity with minimal cracks, voids, wrinkles, ripples, and other contaminants are critical. While this goal has been difficult to achieve, the primary source of defects, contaminants, and inhomogeneity lies in the post-synthesis transfer of 2D materials from the growth substrate onto target substrates. Therefore, direct synthesis of high quality 2D materials on insulating substrates remains a critical technological challenge, although progress is being made, particularly for 2D chalcogenides.[140] Overall, the vast integration possibilities presented by mixed-dimensional vdW heterostructures suggest significant future growth potential for this field in both fundamental studies and applied technologies.

**Acknowledgements**

The authors acknowledge support from the Materials Research Science and Engineering Center (MRSEC) of Northwestern University (NSF DMR-1121262), and the 2-DARE program (NSF EFRI-143510).

**Competing financial interests**

The authors declare no competing financial interests.

**Figures**

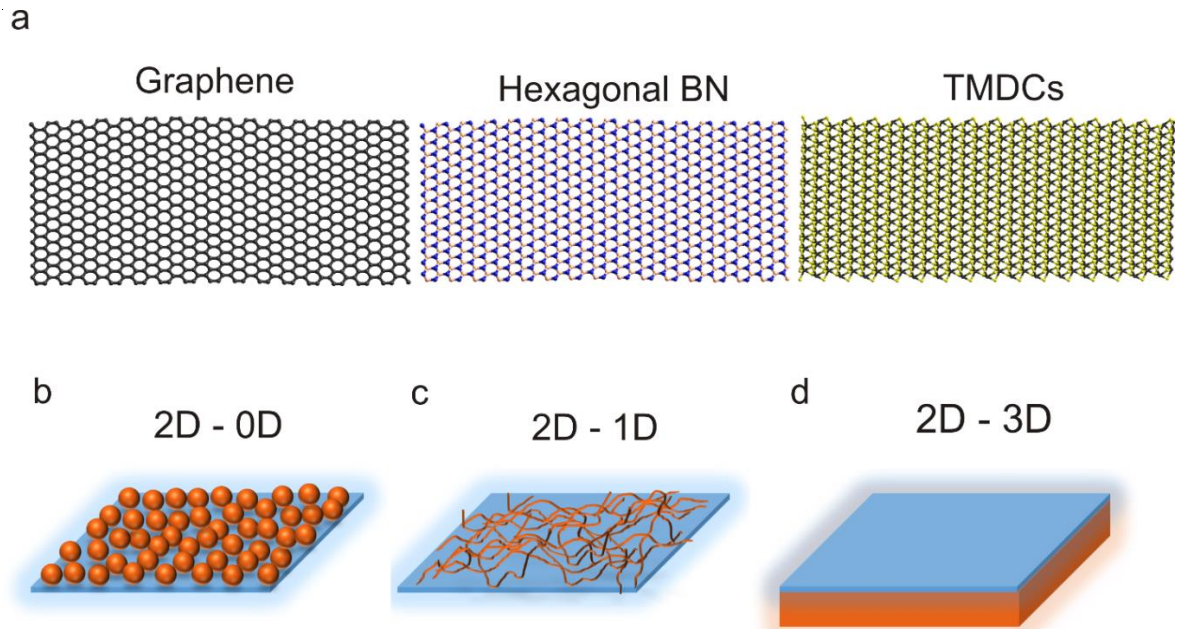

**Figure 1.** Schematic illustrations of prototypical 2D materials and mixed-dimensional van der Waals (vdW) heterojunctions. **a,** Ball and stick models of the three primary 2D materials under consideration: semi-metallic graphene, insulating hexagonal boron nitride (h-BN), and semiconducting transition metal dichalcogenides (TMDCs). **b,** A heterojunction between a 2D material and 0D semiconductors such as quantum dots or small organic molecules. **c,** A heterojunction between a 2D material and 1D materials such as nanotubes, nanowires, and polymers. **d,** A heterojunction between a 2D material and 3D semiconductors such as bulk Si, amorphous oxides, and III-V compound semiconductors.

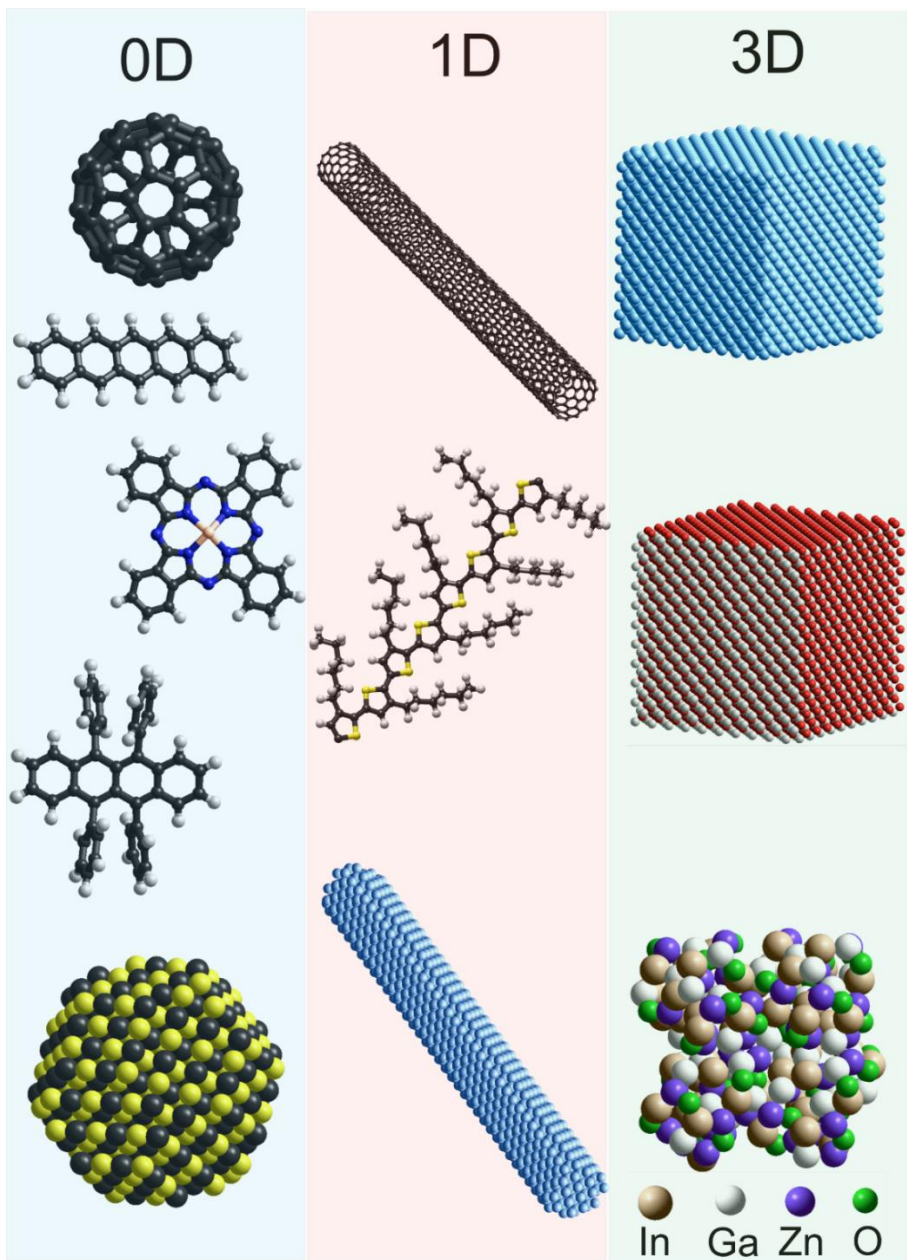

**Figure 2**. Schematic examples of 0D, 1D, and 3D semiconductor materials. 0D (top to bottom): fullerenes ($C_{60}$), small organic molecules (pentacene, copper phthalocyanine, rubrene), and lead sulfide quantum dot. 1D: single-walled carbon nanotubes (SWCNTs), semiconducting conjugated polymers (poly 3-hexathiophene), and Si semiconductor nanowire. 3D: bulk inorganic semiconductors (Si, GaAs) and amorphous oxide semiconductors (IGZO).

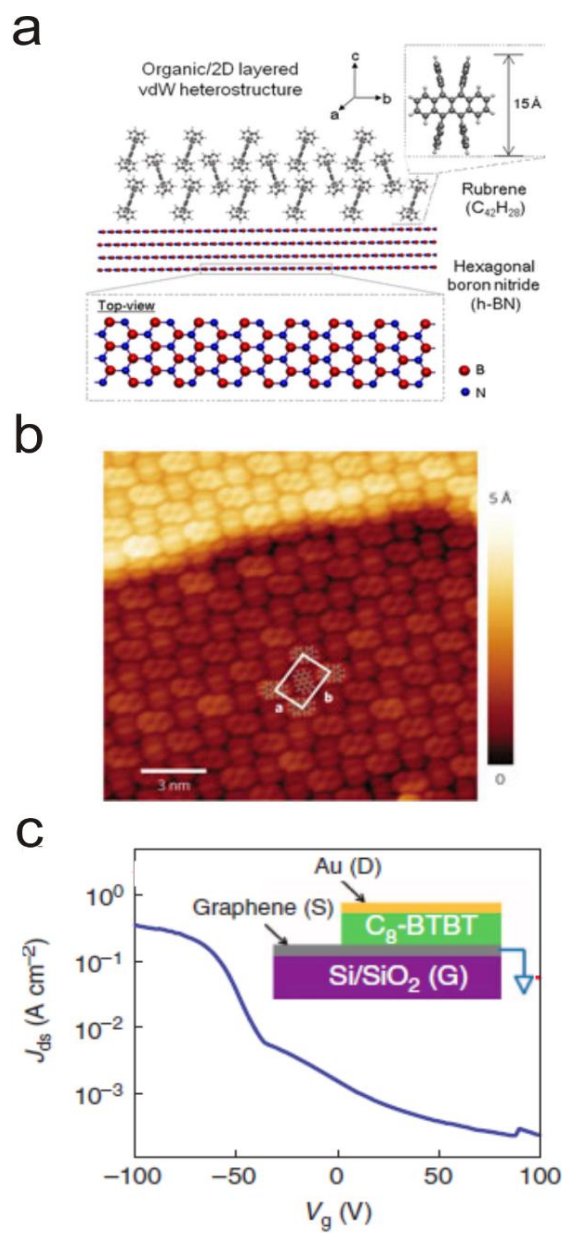

**Figure 3.** Organic-2D heterostructures. **a,** Schematic diagram of rubrene single crystal assembly on h-BN. **b,** Ordered self-assembled monolayer of perylene tetracarboxy dianhydride (PTCDA) on epitaxial graphene with a herringbone structure. PTCDA molecular orientation is 'face-on' due to the π-π interactions between the graphene and PTCDA. **c,** Vertical organic field effect transistor (OFET) schematic based on a heterostructure of 15 nm dioctylbenzothienobenzothiophene (C8-BTBT) self-assembled on graphene (inset). The high on-current density and on-off ratio are apparent in the plot (applied drain-source voltage = 1 V). Figures reproduced with permission from: **a,** ref. [45] © Wiley-VCH. **b,** ref.[30] and **c,** ref.[60] © Macmillan publishers Ltd.

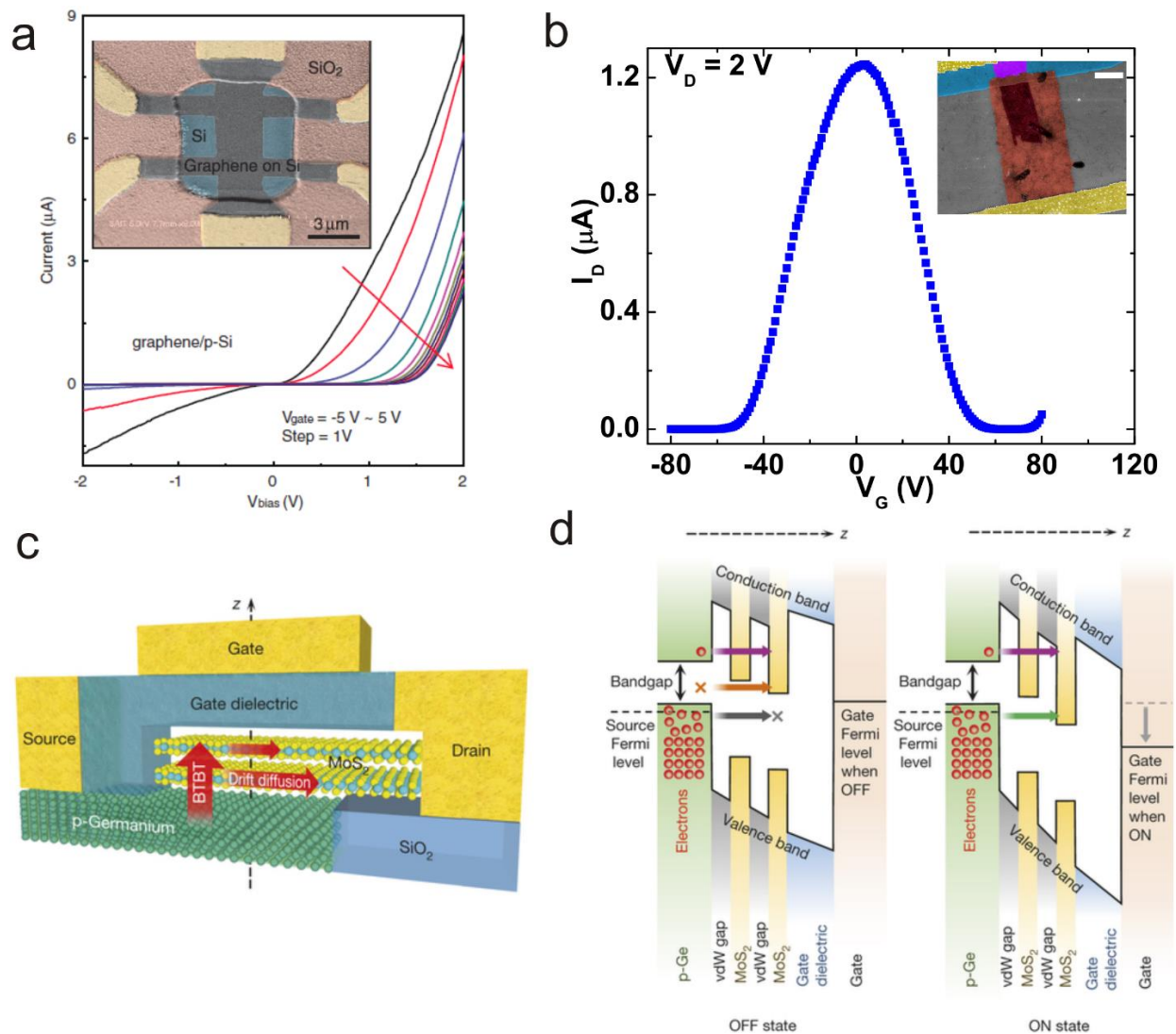

**Figure 4.** Gate-tunable heterojunction devices. **a,** CVD graphene/p-Si, gate-tunable Schottky diode. The strong modulation of the source-drain current as a function of the gate bias is evident at a forward bias of 1 V. Inset shows a scanning electron micrograph of a representative device. **b,** Anti-ambipolar transfer characteristic of a gate-tunable p-n heterojunction based on p-type SWCNTs and n-type $MoS_2$. The transconductance ($dI_D/dV_G$) changes sign at $V_G = 0$ V. Inset shows an SEM micrograph of a representative device (scale bar = 2.5 μm). **c,** Schematic representation of a heavily doped p-Ge/n-$MoS_2$ heterojunction tunneling field-effect transistor. **d,** Band diagram of the device in (c), in (left) off-state and (right) on-state. The tunneling pathway (green arrow)

opens since the MoS$_2$ conduction band is lowered by the applied gate voltage. Figures reproduced with permission from: **a,** ref.[67] © American Association for the Advancement of Science (AAAS). **b,** ref.[34] © National Academy of Sciences, U.S.A. **c-d,** ref.[72] © Macmillan Publishers Ltd.

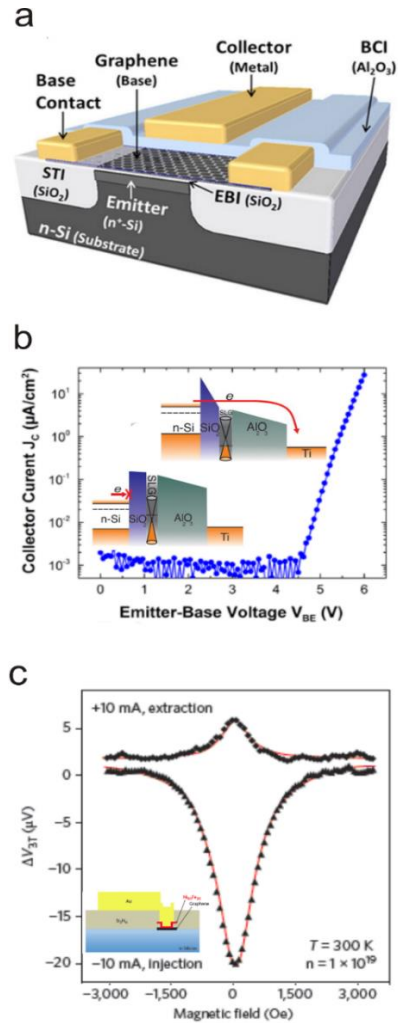

**Figure 5.** Tunneling transport in mixed-dimensional vdW heterostructures. **a,** Schematic diagram of a hot electron transistor (HET) using graphene as a base and n-type Si as the emitter. Carriers tunnel in the vertical direction from the Si emitter to the metal collector via the graphene base and oxide tunnel barriers. The voltage on the graphene base modulates the height of the tunnel barriers and controls the switching function. **b,** Typical collector current versus base voltage (J-V) characteristic of a graphene HET showing collector current on/off ratios in excess 10[4]. Inset shows the band diagrams in the on and off states. **c,** Room-temperature Hanle effect demonstration for spin injection and extraction in a NiFe/graphene/Si heterostructure. Figures reproduced with permission from: **a,** ref.[75] © The American Chemical Society (ACS). **b,** ref.[73, 75] © ACS, Elsevier B.V. **c,** ref.[79] © Macmillan publishers Ltd.

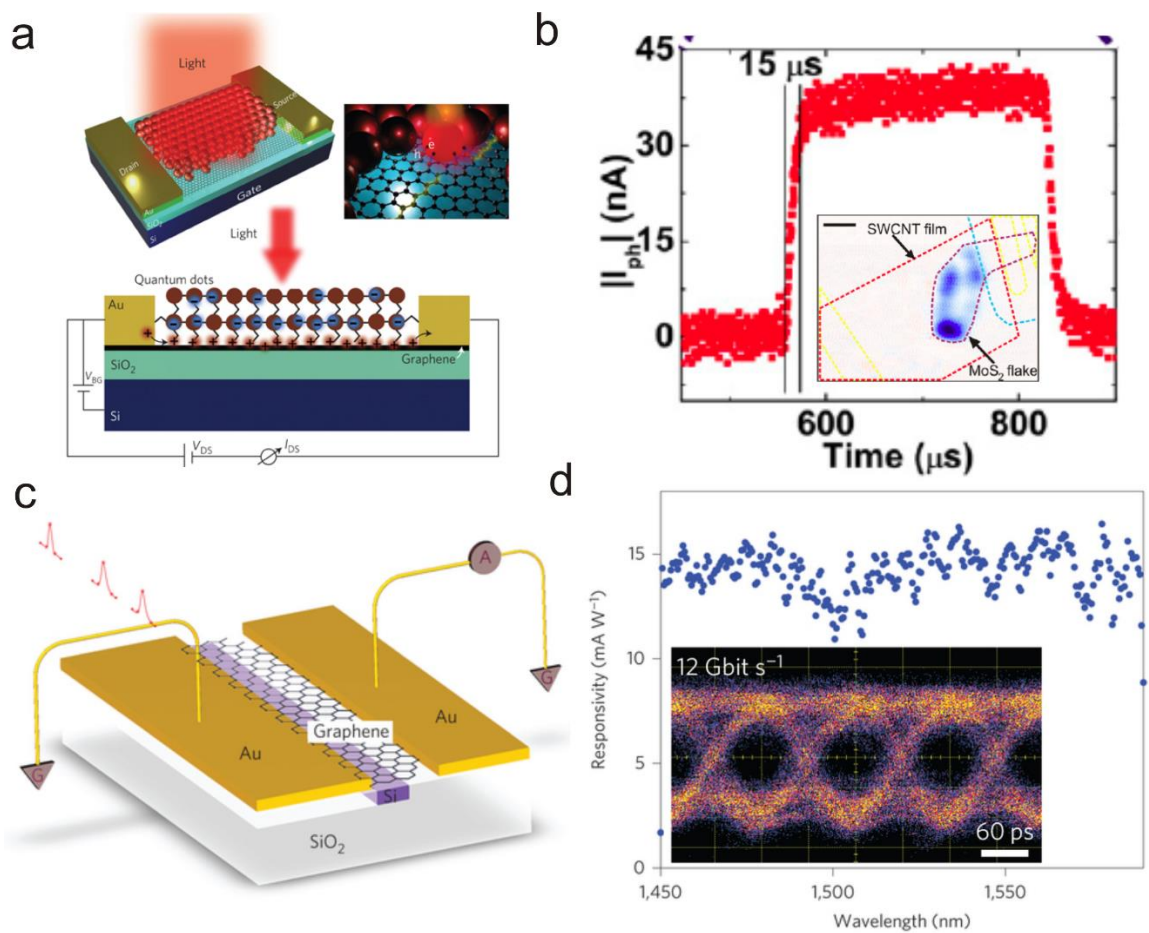

**Figure 6.** Photodetectors based on mixed-dimensional vdW heterostructures. **a,** Schematic representation of a 0D-2D lead sulfide (PbS) quantum dot-sensitized graphene phototransistor. The QDs absorb light and then transfer the photoexcited holes to graphene, thereby doping the graphene and producing the photoresponse. **b,** 1D-2D SWCNT-MoS$_2$ p-n junction photodiode showing a photoresponse of 15 µs. Inset shows a scanning photocurrent micrograph with the photoresponse localized on the overlapping junction region between the two semiconductors (scale bar = 2.5 µm). **c,** Schematic representation of a Si waveguide integrated graphene phototransistor and its measurement setup. **d,** Zero bias photoresponsivity of the phototransistor depicted in (c), showing a broadband response in the near-infrared region. Inset shows a 12 Gbit/s optical data test

link for a representative device, demonstrating the GHz range cut-off frequency. Figures reproduced with permission from: **a,** ref.[32] © Macmillan publishers Ltd.; **b,** ref.[34] © National Academy of Sciences, U.S.A.; **c-d,** ref.[102] © Macmillan publishers Ltd.

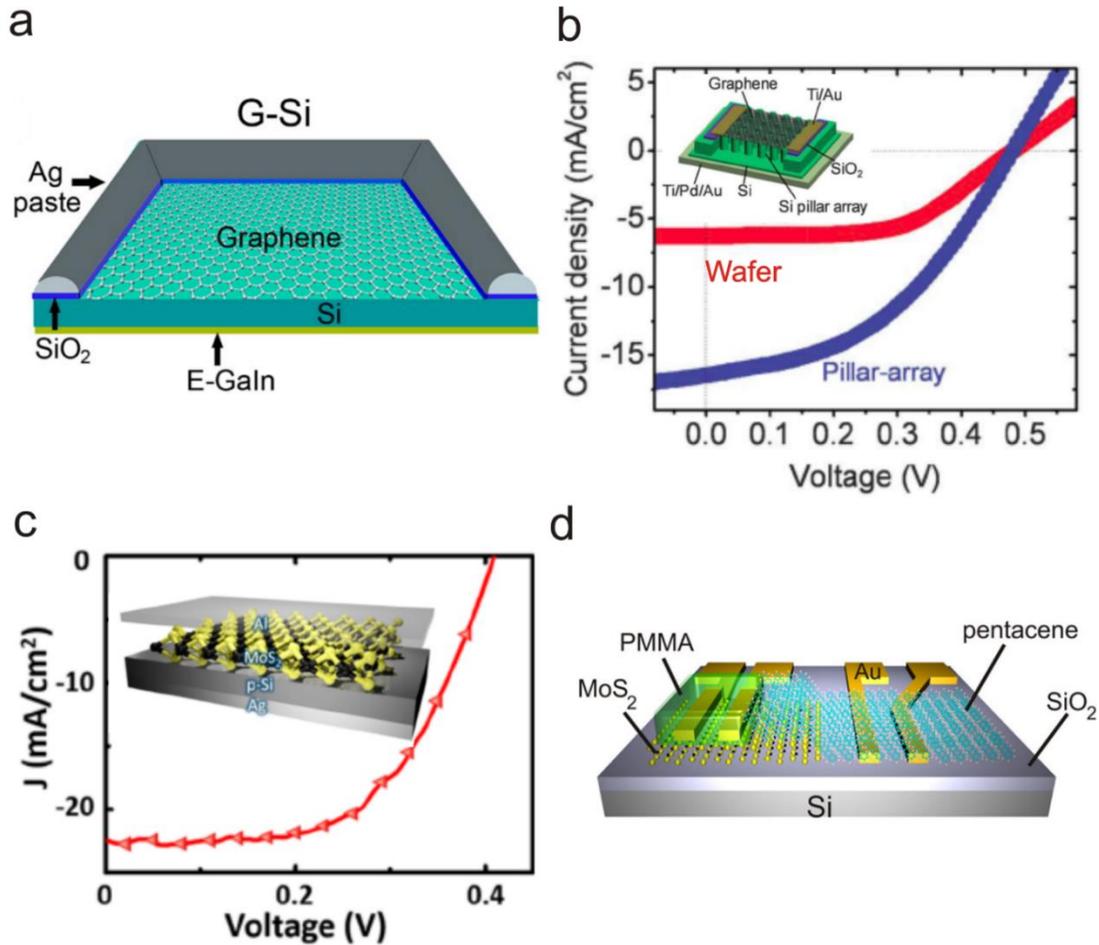

**Figure 7.** Photovoltaics based on mixed-dimensional vdW heterostructures. **a,** Schematic illustration of a graphene-Si heterostructure photovoltaic device showing an Ag front contact and Ga-In alloy back contact. The $SiO_2$ layer between the Ag and Si prevents shorting. **b,** Current-voltage characteristics of a planar graphene-Si heterojunction compared to a graphene-Si micropillar array junction. The higher short circuit current in the case of the Si micropillar array indicates higher carrier collection efficiency. Inset shows a schematic diagram of a Si micropillar array device. **c,** Current-voltage characteristics of a p-Si/n-$MoS_2$ p-n heterojunction device under illumination showing a pronounced photovoltaic effect. Inset shows the device schematic. **d,**

Schematic diagram of a lateral photovoltaic device based on n-MoS2/p-pentacene. The lateral device geometry combined with the underlying gate allows gate-tunability in this device structure. Figures reproduced with permission from: **a,** ref.[107] © ACS; **b,**[109] © Royal Society of Chemistry (RSC); **c,** ref.[121]; **d,** ref.[69] © ACS.

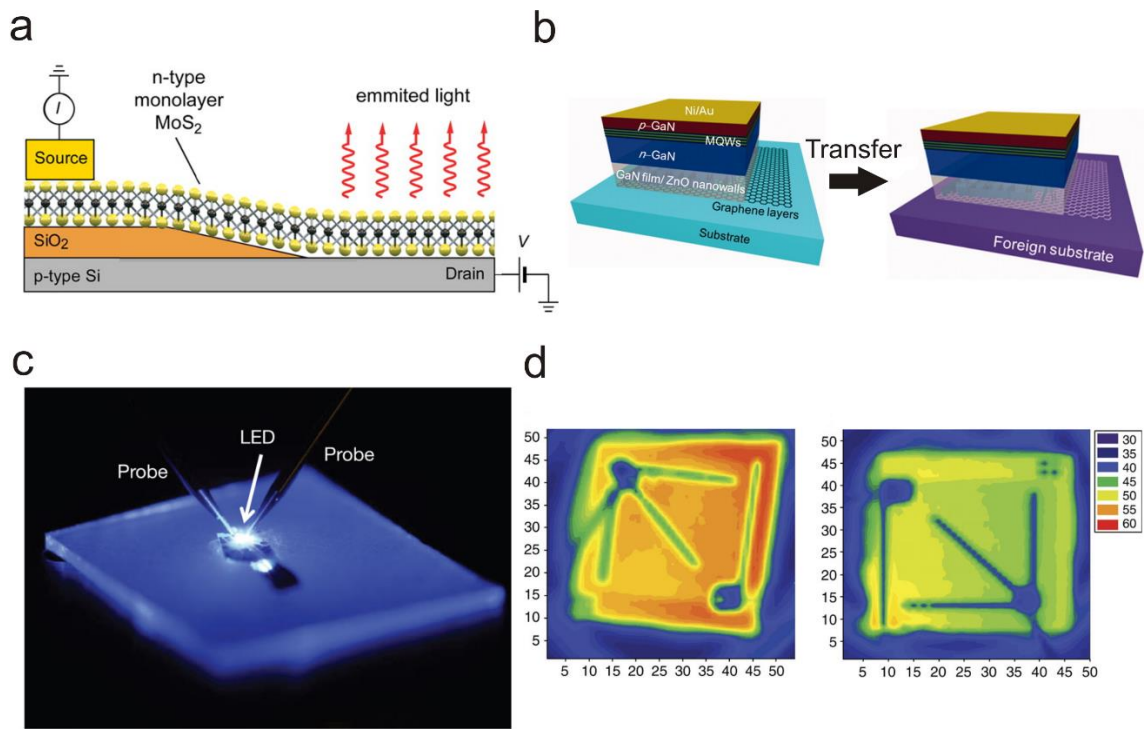

**Figure 8.** Light-emitting heterostructure devices. **a,** Schematic diagram of a monolayer n-MoS2/p-Si heterojunction diode that emits light under forward bias. Since Si has an indirect band gap, the emission is only from MoS2 which originates from radiative recombination of excitons. **b,** Structure and transfer scheme of a GaN/InGaN multiple quantum well (MQW) light-emitting diode (LED) grown on a graphene buffer layer. **c,** Photograph of a GaN/InGaN MQW LED under operation grown on a h-BN buffer layer transferred onto an arbitrary substrate. **d,** Surface temperature distributions of conventional (left) and embedded graphene layer (right) GaN LEDs under 100 mA current injection. The lower surface temperatures on the graphene-embedded LED is evident as a result of better heat dissipation by the graphene. Figures reproduced with permission from: **a,** ref.[120] © ACS. **b,** ref.[37] © AAAS. **c,** ref.[38]; **d,** ref.[134] © Macmillan publishers Ltd.

# References

1. Novoselov K. S., Geim A. K., Morozov S., Jiang D., Zhang Y., Dubonos S.*, et al.* Electric field effect in atomically thin carbon films. *Science*, **306**, 666-669 (2004).

2. Jariwala D., Sangwan V. K., Lauhon L. J., Marks T. J., Hersam M. C. Emerging device applications for semiconducting two-dimensional transition metal dichalcogenides. *ACS Nano*, **8**, 1102–1120 (2014).

3. Britnell L., Gorbachev R. V., Jalil R., Belle B. D., Schedin F., Mishchenko A.*, et al.* Field-effect tunneling transistor based on vertical graphene heterostructures. *Science*, **335**, 947-950 (2012).

4. Britnell L., Ribeiro R. M., Eckmann A., Jalil R., Belle B. D., Mishchenko A.*, et al.* Strong light-matter interactions in heterostructures of atomically thin films. *Science*, **340**, 1311-1314 (2013).

5. Georgiou T., Jalil R., Belle B. D., Britnell L., Gorbachev R. V., Morozov S. V.*, et al.* Vertical field-effect transistor based on graphene-$WS_2$ heterostructures for flexible and transparent electronics. *Nat Nanotechnol*, **8**, 100-103 (2013).

6. Mishchenko A., Tu J. S., CaoY, Gorbachev R. V., Wallbank J. R., Greenaway M. T.*, et al.* Twist-controlled resonant tunnelling in graphene/boron nitride/graphene heterostructures. *Nat Nanotechnol*, **9**, 808-813 (2014).

7. Yu W. J., Li Z., Zhou H., Chen Y., Wang Y., Huang Y.*, et al.* Vertically stacked multi-heterostructures of layered materials for logic transistors and complementary inverters. *Nat Mater*, **12**, 246-252 (2012).

8. Yu W. J., Liu Y., Zhou H., Yin A., Li Z., Huang Y.*, et al.* Highly efficient gate-tunable photocurrent generation in vertical heterostructures of layered materials. *Nat Nanotechnol*, **8**, 952–958 (2013).

9. Grigorieva I. V., Geim A. K. Van der Waals heterostructures. *Nature*, **499**, 419-425 (2013).

10. Das S., Robinson J. A., Dubey M., Terrones H., Terrones M. Beyond Graphene: Progress in Novel Two-Dimensional Materials and van der Waals Solids. *Annu Rev Mater Res*, **45**, 1-27 (2015).

11. Wang H., Yuan H., Sae Hong S., Li Y., Cui Y. Physical and chemical tuning of two-dimensional transition metal dichalcogenides. *Chem Soc Rev*, **44**, 2664-2680 (2015).

12.    Chhowalla M., Shin H. S., Eda G., Li L.-J., Loh K. P., Zhang H. The Chemistry of Two-Dimensional Layered Transition Metal Dichalcogenide Nanosheets. *Nat Chem*, **5**, 263-275 (2013).

13.    Hoppe H., Sariciftci N. S. Organic solar cells: An overview. *J Mater Res*, **19**, 1924-1945 (2004).

14.    Gong M., Shastry T. A., Xie Y., Bernardi M., Jasion D., Luck K. A.*, et al.* Polychiral Semiconducting Carbon Nanotube–Fullerene Solar Cells. *Nano Lett*, **14**, 5308-5314 (2014).

15.    Neto A. C., Guinea F., Peres N., Novoselov K. S., Geim A. K. The Electronic Properties of Graphene. *Rev Mod Phys*, **81**, 109-162 (2009).

16.    Wang Q. H., Kalantar-Zadeh K., Kis A., Coleman J. N., Strano M. S. Electronics and optoelectronics of two-dimensional transition metal dichalcogenides. *Nat Nanotechnol*, **7**, 699-712 (2012).

17.    Kroto H. W., Heath J. R., O'Brien S. C., Curl R. F., Smalley R. E. C60: Buckminsterfullerene. *Nature*, **318**, 162-163 (1985).

18.    Facchetti A. Semiconductors for organic transistors. *Mater Today*, **10**, 28-37 (2007).

19.    Alivisatos A. P. Semiconductor Clusters, Nanocrystals, and Quantum Dots. *Science*, **271**, 933-937 (1996).

20.    Arnold M. S., Green A. A., Hulvat J. F., Stupp S. I., Hersam M. C. Sorting carbon nanotubes by electronic structure using density differentiation. *Nat Nanotechnol*, **1**, 60-65 (2006).

21.    Lieber C. M., Wang Z. L. Functional Nanowires. *MRS Bull*, **32**, 99-108 (2007).

22.    Yu X., Marks T. J., Facchetti A. Metal oxides for optoelectronic applications. *Nat Mater*, **15**, 383-396 (2016).

23.    Li S.-L., Tsukagoshi K., Orgiu E., Samorì P. Charge transport and mobility engineering in two-dimensional transition metal chalcogenide semiconductors. *Chem Soc Rev*, **45**, 118-151 (2016).

24.    Kang J., Liu W., Sarkar D., Jena D., Banerjee K. Computational study of metal contacts to monolayer transition-metal dichalcogenide semiconductors. *Phys Rev X*, **4**, 031005 (2014).

25.     Allain A., Kang J., Banerjee K., Kis A. Electrical contacts to two-dimensional semiconductors. *Nat Mater*, **14**, 1195-1205 (2015).

26.     Sze S. M., Ng K. K. *Physics of Semiconductor Devices*, 3rd edn. Wiley, 2007.

27.     Ishii H., Sugiyama K., Ito E., Seki K. Energy Level Alignment and Interfacial Electronic Structures at Organic/Metal and Organic/Organic Interfaces. *Adv Mater*, **11**, 605-625 (1999).

28.     So F. *Organic electronics: materials, processing, devices and applications*. CRC press, 2009.

29.     Jariwala D., Sangwan V. K., Lauhon L. J., Marks T. J., Hersam M. C. Carbon nanomaterials for electronics, optoelectronics, photovoltaics, and sensing. *Chem Soc Rev*, **42**, 2824-2860 (2013).

30.     Wang Q. H., Hersam M. C. Room-temperature molecular-resolution characterization of self-assembled organic monolayers on epitaxial graphene. *Nat Chem*, **1**, 206-211 (2009).

31.     Kufer D., Nikitskiy I., Lasanta T., Navickaite G., Koppens F. H. L., Konstantatos G. Hybrid 2D–0D MoS2–PbS Quantum Dot Photodetectors. *Adv Mater*, **27**, 176-180 (2015).

32.     Konstantatos G., Badioli M., Gaudreau L., Osmond J., Bernechea M., de Arquer F. P. G.*, et al.* Hybrid graphene-quantum dot phototransistors with ultrahigh gain. *Nat Nanotechnol*, **7**, 363-368 (2012).

33.     Jariwala D., Sangwan V. K., Seo J.-W. T., Xu W., Smith J., Kim C. H.*, et al.* Large-Area, Low-Voltage, Antiambipolar Heterojunctions from Solution-Processed Semiconductors. *Nano Lett*, **15**, 416-421 (2015).

34.     Jariwala D., Sangwan V. K., Wu C.-C., Prabhumirashi P. L., Geier M. L., Marks T. J.*, et al.* Gate-Tunable Carbon Nanotube–MoS$_2$ Heterojunction p-n Diode. *Proc Nat Acad Sci USA*, **110**, 18076–18080 (2013).

35.     Lee J.-H., Lee E. K., Joo W.-J., Jang Y., Kim B.-S., Lim J. Y.*, et al.* Wafer-Scale Growth of Single-Crystal Monolayer Graphene on Reusable Hydrogen-Terminated Germanium. *Science*, **344**, 286-289 (2014).

36.     Ruzmetov D., Zhang K., Stan G., Kalanyan B., Bhimanapati G. R., Eichfeld S. M.*, et al.* Vertical 2D/3D Semiconductor Heterostructures Based on Epitaxial Molybdenum Disulfide and Gallium Nitride. *ACS Nano*, **10**, 3580-3588 (2016).

37.     Chung K., Lee C.-H., Yi G.-C. Transferable GaN Layers Grown on ZnO-Coated Graphene Layers for Optoelectronic Devices. *Science*, **330**, 655-657 (2010).

38.     Kobayashi Y., Kumakura K., Akasaka T., Makimoto T. Layered boron nitride as a release layer for mechanical transfer of GaN-based devices. *Nature*, **484**, 223-227 (2012).

39.     Kory M. J., Wörle M., Weber T., Payamyar P., van de PollStan W., Dshemuchadse J.*, et al.* Gram-scale synthesis of two-dimensional polymer crystals and their structure analysis by X-ray diffraction. *Nat Chem*, **6**, 779-784 (2014).

40.     Pfeffermann M., Dong R., Graf R., Zajaczkowski W., Gorelik T., Pisula W.*, et al.* Free-Standing Monolayer Two-Dimensional Supramolecular Organic Framework with Good Internal Order. *J Am Chem Soc*, **137**, 14525-14532 (2015).

41.     Kissel P., Murray D. J., Wulftange W. J., Catalano V. J., King B. T. A nanoporous two-dimensional polymer by single-crystal-to-single-crystal photopolymerization. *Nat Chem*, **6**, 774-778 (2014).

42.     Zhuang X., Gehrig D., Forler N., Liang H., Wagner M., Hansen M. R.*, et al.* Conjugated Microporous Polymers with Dimensionality-Controlled Heterostructures for Green Energy Devices. *Adv Mater*, **27**, 3789-3796 (2015).

43.     Murray D. J., Patterson D. D., Payamyar P., Bhola R., Song W., Lackinger M.*, et al.* Large Area Synthesis of a Nanoporous Two-Dimensional Polymer at the Air/Water Interface. *J Am Chem Soc*, **137**, 3450-3453 (2015).

44.     Tour J. M. Molecular Electronics. Synthesis and Testing of Components. *Acc Chem Res*, **33**, 791-804 (2000).

45.     Lee C.-H., Schiros T., Santos E. J. G., Kim B., Yager K. G., Kang S. J.*, et al.* Epitaxial Growth of Molecular Crystals on van der Waals Substrates for High-Performance Organic Electronics. *Adv Mater*, **26**, 2812-2817 (2014).

46.     Kang S. J., Lee G.-H., Yu Y.-J., Zhao Y., Kim B., Watanabe K.*, et al.* Organic Field Effect Transistors Based on Graphene and Hexagonal Boron Nitride Heterostructures. *Adv Funct Mater*, **24**, 5157-5163 (2014).

47.     Lee G.-H., Lee C.-H., van der Zande A. M., Han M., Cui X., Arefe G.*, et al.* Heterostructures based on inorganic and organic van der Waals systems. *APL Mat*, **2**, 092511 (2014).

48.     Jo S. B., Kim H. H., Lee H., Kang B., Lee S., Sim M.*, et al.* Boosting Photon Harvesting in Organic Solar Cells with Highly Oriented Molecular Crystals via Graphene–Organic Heterointerface. *ACS Nano*, **9**, 8206-8219 (2015).

49.     Lee W. H., Park J., Sim S. H., Lim S., Kim K. S., Hong B. H.*, et al.* Surface-Directed Molecular Assembly of Pentacene on Monolayer Graphene for High-Performance Organic Transistors. *J Am Chem Soc*, **133**, 4447-4454 (2011).

50.     Lee S., Kang S.-J., Jo G., Choe M., Park W., Yoon J.*, et al.* Enhanced characteristics of pentacene field-effect transistors with graphene electrodes and substrate treatments. *Appl Phys Lett*, **99**, 083306 (2011).

51.     Lee S., Jo G., Kang S.-J., Wang G., Choe M., Park W.*, et al.* Enhanced Charge Injection in Pentacene Field-Effect Transistors with Graphene Electrodes. *Adv Mater*, **23**, 100-105 (2011).

52.     Basu S., Lee M. C., Wang Y.-H. Graphene-based electrodes for enhanced organic thin film transistors based on pentacene. *Phys Chem Chem Phys*, **16**, 16701-16710 (2014).

53.     Di C.-a., Wei D., Yu G., Liu Y., Guo Y., Zhu D. Patterned Graphene as Source/Drain Electrodes for Bottom-Contact Organic Field-Effect Transistors. *Adv Mater*, **20**, 3289-3293 (2008).

54.     Wang Y., Torres J. A., Stieg A. Z., Jiang S., Yeung M. T., Rubin Y.*, et al.* Graphene-Assisted Solution Growth of Vertically Oriented Organic Semiconducting Single Crystals. *ACS Nano*, **9**, 9486-9496 (2015).

55.     Colson J. W., Woll A. R., Mukherjee A., Levendorf M. P., Spitler E. L., Shields V. B.*, et al.* Oriented 2D Covalent Organic Framework Thin Films on Single-Layer Graphene. *Science*, **332**, 228-231 (2011).

56.     Huafeng Y., Freddie W., Elias G., Edward L., Liam B., Alexandre F.*, et al.* Dielectric nanosheets made by liquid-phase exfoliation in water and their use in graphene-based electronics. *2D Mater*, **1**, 011012 (2014).

57.     Schlierf A., Yang H., Gebremedhn E., Treossi E., Ortolani L., Chen L.*, et al.* Nanoscale insight into the exfoliation mechanism of graphene with organic dyes: effect of charge, dipole and molecular structure. *Nanoscale*, **5**, 4205-4216 (2013).

58.     Sarbani B., Feri A., Yeong-Her W. Blending effect of 6,13-bis(triisopropylsilylethynyl) pentacene–graphene composite layers for flexible thin film transistors with a polymer gate dielectric. *Nanotechnology*, **25**, 085201 (2014).

59.     Zhang Y., Liu S., Liu W., Liang T., Yang X., Xu M.*, et al.* Two-dimensional MoS2-assisted immediate aggregation of poly-3-hexylthiophene with high mobility. *Phys Chem Chem Phys*, **17**, 27565-27572 (2015).

60.     He D., Zhang Y., Wu Q., Xu R., Nan H., Liu J.*, et al.* Two-dimensional quasi-freestanding molecular crystals for high-performance organic field-effect transistors. *Nat Commun*, **5**, 5162 (2014).

61.     Parui S., Pietrobon L., Ciudad D., Vélez S., Sun X., Casanova F.*, et al.* Gate-Controlled Energy Barrier at a Graphene/Molecular Semiconductor Junction. *Adv Funct Mater*, **25**, 2972-2979 (2015).

62.     Liu Y., Zhou H., Weiss N. O., Huang Y., Duan X. High-performance organic vertical thin film transistor using graphene as a tunable contact. *ACS Nano*, **9**, 11102-11108 (2015).

63.     Hlaing H., Kim C.-H., Carta F., Nam C.-Y., Barton R. A., Petrone N.*, et al.* Low-Voltage Organic Electronics Based on a Gate-Tunable Injection Barrier in Vertical graphene-organic Semiconductor Heterostructures. *Nano Lett*, **15**, 69-74 (2015).

64.     Nomura K., Ohta H., Takagi A., Kamiya T., Hirano M., Hosono H. Room-Temperature Fabrication of Transparent Flexible Thin-Film Transistors Using Amorphous Oxide Semiconductors. *Nature*, **432**, 488-492 (2004).

65.     Liu Y., Zhou H., Cheng R., Yu W., Huang Y., Duan X. Highly flexible electronics from scalable vertical thin film transistors. *Nano Lett*, **14**, 1413-1418 (2014).

66.     Heo J., Byun K.-E., Lee J., Chung H.-J., Jeon S., Park S.*, et al.* Graphene and thin-film semiconductor heterojunction transistors integrated on wafer scale for low-power electronics. *Nano Lett*, **13**, 5967-5971 (2013).

67.     Yang H., Heo J., Park S., Song H. J., Seo D. H., Byun K.-E.*, et al.* Graphene Barristor, A Triode Device with a Gate-Controlled Schottky Barrier. *Science*, **336**, 1140-1143 (2012).

68.     Ojeda-Aristizabal C., Bao W., Fuhrer M. S. Thin-film barristor: A gate-tunable vertical graphene-pentacene device. *Phys Rev B*, **88**, 035435 (2013).

69.     Jariwala D., Howell S. L., Chen K.-S., Kang J., Sangwan V. K., Filippone S. A.*, et al.* Hybrid, gate-tunable, van der Waals p-n heterojunctions from pentacene and MoS$_2$. *Nano Lett*, **16**, 497–503 (2015).

70.     Velez S., Ciudad D., Island J., Buscema M., Txoperena O., Parui S., *et al.* Gate-tunable diode and photovoltaic effect in an organic-2D layered material p-n junction. *Nanoscale*, (2015).

71.     Jeong H., Bang S., Oh H. M., Jeong H. J., An S.-J., Han G. H., *et al.* Semiconductor–Insulator–Semiconductor Diode Consisting of Monolayer MoS2, h-BN, and GaN Heterostructure. *ACS Nano*, **9**, 10032-10038 (2015).

72.     Sarkar D., Xie X., Liu W., Cao W., Kang J., Gong Y., *et al.* A subthermionic tunnel field-effect transistor with an atomically thin channel. *Nature*, **526**, 91-95 (2015).

73.     Vaziri S., Smith A. D., Östling M., Lupina G., Dabrowski J., Lippert G., *et al.* Going ballistic: Graphene hot electron transistors. *Solid State Commun*, **224**, 64-75 (2015).

74.     Mehr W., Dabrowski J., Christoph Scheytt J., Lippert G., Xie Y.-H., Lemme M. C., *et al.* Vertical Graphene Base Transistor. *IEEE Electron Device Lett*, **33**, 691-693 (2012).

75.     Vaziri S., Lupina G., Henkel C., Smith A. D., Östling M., Dabrowski J., *et al.* A Graphene-Based Hot Electron Transistor. *Nano Lett*, **13**, 1435-1439 (2013).

76.     Torres C. M., Lan Y.-W., Zeng C., Chen J.-H., Kou X., Navabi A., *et al.* High-Current Gain Two-Dimensional MoS2-Base Hot-Electron Transistors. *Nano Lett*, **15**, 7905-7912 (2015).

77.     Vaziri S., Belete M., Dentoni Litta E., Smith A. D., Lupina G., Lemme M. C., *et al.* Bilayer insulator tunnel barriers for graphene-based vertical hot-electron transistors. *Nanoscale*, **7**, 13096-13104 (2015).

78.     Mead C. A. Operation of Tunnel-Emission Devices. *J Appl Phys*, **32**, 646-652 (1961).

79.     van 't Erve O. M. J., Friedman A. L., CobasE, Li C. H., Robinson J. T., Jonker B. T. Low-resistance spin injection into silicon using graphene tunnel barriers. *Nat Nanotechnol*, **7**, 737-742 (2012).

80.     van 't Erve O. M. J., Friedman A. L., Li C. H., Robinson J. T., Connell J., Lauhon L. J., *et al.* Spin transport and Hanle effect in silicon nanowires using graphene tunnel barriers. *Nat Commun*, **6**, (2015).

81.     Koppens F. H. L., Mueller T., Avouris P., Ferrari A. C., Vitiello M. S., Polini M. Photodetectors based on graphene, other two-dimensional materials and hybrid systems. *Nat Nanotechnol*, **9**, 780-793 (2014).

82.   Sun Z., Liu Z., Li J., Tai G.-a., Lau S.-P., Yan F. Infrared Photodetectors Based on CVD-Grown Graphene and PbS Quantum Dots with Ultrahigh Responsivity. *Adv Mater*, **24**, 5878-5883 (2012).

83.   Jang S., Hwang E., Lee Y., Lee S., Cho J. H. Multifunctional Graphene Optoelectronic Devices Capable of Detecting and Storing Photonic Signals. *Nano Lett*, **15**, 2542-2547 (2015).

84.   Roy K., Padmanabhan M., Goswami S., Sai T. P., Ramalingam G., Raghavan S.*, et al.* Graphene-$MoS_2$ Hybrid Structures for Multifunctional Photoresponsive Memory Devices. *Nat Nanotechnol*, **8**, 826–830 (2013).

85.   Yu S. H., Lee Y., Jang S. K., Kang J., Jeon J., Lee C.*, et al.* Dye-Sensitized MoS2 Photodetector with Enhanced Spectral Photoresponse. *ACS Nano*, **8**, 8285-8291 (2014).

86.   Cho E., Song W., Park C., Kim J., Kim S., Joo J. Enhancement of photoresponsive electrical characteristics of multilayer MoS2 transistors using rubrene patches. *Nano Res*, **8**, 790-800 (2015).

87.   An X., Liu F., Jung Y. J., Kar S. Tunable Graphene–Silicon Heterojunctions for Ultrasensitive Photodetection. *Nano Lett*, **13**, 909-916 (2013).

88.   Zeng L.-H., Wang M.-Z., Hu H., Nie B., Yu Y.-Q., Wu C.-Y.*, et al.* Monolayer Graphene/Germanium Schottky Junction As High-Performance Self-Driven Infrared Light Photodetector. *ACS Appl Mater Interfaces*, **5**, 9362-9366 (2013).

89.   Chen C.-C., Aykol M., Chang C.-C., Levi A. F. J., Cronin S. B. Graphene-Silicon Schottky Diodes. *Nano Lett*, **11**, 1863-1867 (2011).

90.   Zhu M., Li X., Guo Y., Li X., Sun P., Zang X.*, et al.* Vertical junction photodetectors based on reduced graphene oxide/silicon Schottky diodes. *Nanoscale*, **6**, 4909-4914 (2014).

91.   An Y., Behnam A., Pop E., Ural A. Metal-semiconductor-metal photodetectors based on graphene/p-type silicon Schottky junctions. *Appl Phys Lett*, **102**, 013110 (2013).

92.   Nie B., Hu J. G., Luo L. B., Xie C., Zeng L. H., Lv P.*, et al.* Monolayer Graphene Film on ZnO Nanorod Array for High-Performance Schottky Junction Ultraviolet Photodetectors. *Small*, **9**, 2872-2879 (2013).

93.   Gao Z., Jin W., Zhou Y., Dai Y., Yu B., Liu C.*, et al.* Self-powered flexible and transparent photovoltaic detectors based on CdSe nanobelt/graphene Schottky junctions. *Nanoscale*, **5**, 5576-5581 (2013).

94. Shin D. H., Kim S., Kim J. M., Jang C. W., Kim J. H., Lee K. W.*, et al.* Graphene/Si-Quantum-Dot Heterojunction Diodes Showing High Photosensitivity Compatible with Quantum Confinement Effect. *Adv Mater*, **27**, 2614-2620 (2015).

95. Miao J., Hu W., Guo N., Lu Z., Liu X., Liao L.*, et al.* High-Responsivity Graphene/InAs Nanowire Heterojunction Near-Infrared Photodetectors with Distinct Photocurrent On/Off Ratios. *Small*, **11**, 936-942 (2015).

96. Esmaeili-Rad M. R., Salahuddin S. High Performance Molybdenum Disulfide Amorphous Silicon Heterojunction Photodetector. *Sci Rep*, **3**, 2345 (2013).

97. Liu F., Chow W. L., He X., Hu P., Zheng S., Wang X.*, et al.* Van der Waals p–n Junction Based on an Organic–Inorganic Heterostructure. *Adv Funct Mater*, 10.1002/adfm.201502316 (2016).

98. Koester S. J., Li M. Waveguide-coupled graphene optoelectronics. *IEEE J Sel Top Quantum Electron*, **20**, 84-94 (2014).

99. Furchi M., Urich A., Pospischil A., Lilley G., Unterrainer K., Detz H.*, et al.* Microcavity-Integrated Graphene Photodetector. *Nano Lett*, **12**, 2773-2777 (2012).

100. Liu M., Yin X., Ulin-Avila E., Geng B., Zentgraf T., Ju L.*, et al.* A graphene-based broadband optical modulator. *Nature*, **474**, 64-67 (2011).

101. Pospischil A., Humer M., Furchi M. M., Bachmann D., Guider R., Fromherz T.*, et al.* CMOS-compatible graphene photodetector covering all optical communication bands. *Nat Photon*, **7**, 892-896 (2013).

102. Gan X., Shiue R.-J., Gao Y., Meric I., Heinz T. F., Shepard K.*, et al.* Chip-integrated ultrafast graphene photodetector with high responsivity. *Nat Photon*, **7**, 883-887 (2013).

103. Shiue R.-J., Gao Y., Wang Y., Peng C., Robertson A. D., Efetov D. K.*, et al.* High-Responsivity Graphene–Boron Nitride Photodetector and Autocorrelator in a Silicon Photonic Integrated Circuit. *Nano Lett*, **15**, 7288-7293 (2015).

104. Wang X., Cheng Z., Xu K., Tsang H. K., Xu J.-B. High-responsivity graphene/silicon-heterostructure waveguide photodetectors. *Nat Photon*, **7**, 888-891 (2013).

105. Li X., Zhu H., Wang K., Cao A., Wei J., Li C.*, et al.* Graphene-On-Silicon Schottky Junction Solar Cells. *Adv Mater*, **22**, 2743-2748 (2010).

106. Brus V. V., Gluba M. A., Zhang X., Hinrichs K., Rappich J., Nickel N. H. Stability of graphene–silicon heterostructure solar cells. *Phys Status Solidi A*, **211**, 843-847 (2014).

107. Shi E., Li H., Yang L., Zhang L., Li Z., Li P., *et al.* Colloidal Antireflection Coating Improves Graphene–Silicon Solar Cells. *Nano Lett*, **13**, 1776-1781 (2013).

108. Miao X., Tongay S., Petterson M. K., Berke K., Rinzler A. G., Appleton B. R., *et al.* High Efficiency Graphene Solar Cells by Chemical Doping. *Nano Lett*, **12**, 2745-2750 (2012).

109. Lin Y., Li X., Xie D., Feng T., Chen Y., Song R., *et al.* Graphene/semiconductor heterojunction solar cells with modulated antireflection and graphene work function. *Energ Environ Sci*, **6**, 108-115 (2013).

110. Zhang X., Xie C., Jie J., Zhang X., Wu Y., Zhang W. High-efficiency graphene/Si nanoarray Schottky junction solar cells via surface modification and graphene doping. *J Mater Chem A*, **1**, 6593-6601 (2013).

111. Feng T., Xie D., Lin Y., Zang Y., Ren T., Song R., *et al.* Graphene based Schottky junction solar cells on patterned silicon-pillar-array substrate. *Appl Phys Lett*, **99**, 233505 (2011).

112. Xie C., Lv P., Nie B., Jie J., Zhang X., Wang Z., *et al.* Monolayer graphene film/silicon nanowire array Schottky junction solar cells. *Appl Phys Lett*, **99**, 133113 (2011).

113. Song Y., Li X., Mackin C., Zhang X., Fang W., Palacios T., *et al.* Role of Interfacial Oxide in High-Efficiency Graphene–Silicon Schottky Barrier Solar Cells. *Nano Lett*, **15**, 2104-2110 (2015).

114. Li X., Chen W., Zhang S., Wu Z., Wang P., Xu Z., *et al.* 18.5% efficient graphene/GaAs van der Waals heterostructure solar cell. *Nano Energy*, **16**, 310-319 (2015).

115. Vazquez-Mena O., Bosco J. P., Ergen O., Rasool H. I., Fathalizadeh A., Tosun M., *et al.* Performance Enhancement of a Graphene-Zinc Phosphide Solar Cell Using the Electric Field-Effect. *Nano Lett*, **14**, 4280-4285 (2014).

116. Zhang L., Fan L., Li Z., Shi E., Li X., Li H., *et al.* Graphene-CdSe nanobelt solar cells with tunable configurations. *Nano Res*, **4**, 891-900 (2011).

117. Ye Y., Gan L., Dai L., Dai Y., Guo X., Meng H., *et al.* A simple and scalable graphene patterning method and its application in CdSe nanobelt/graphene Schottky junction solar cells. *Nanoscale*, **3**, 1477-1481 (2011).

118. Lin S., Li X., Zhang S., Wang P., Xu Z., Zhong H., *et al.* Graphene/CdTe heterostructure solar cell and its enhancement with photo-induced doping. *Appl Phys Lett*, **107**, 191106 (2015).

119. Ye Y., Dai L. Graphene-based Schottky junction solar cells. *J Mater Chem*, **22**, 24224-24229 (2012).

120. Lopez-Sanchez O., Alarcon Llado E., Koman V., Fontcuberta i Morral A., Radenovic A., Kis A. Light Generation and Harvesting in a van der Waals Heterostructure. *ACS Nano*, **8**, 3042-3048 (2014).

121. Tsai M.-L., Su S.-H., Chang J.-K., Tsai D.-S., Chen C.-H., Wu C.-I., *et al.* Monolayer MoS2 Heterojunction Solar Cells. *ACS Nano*, **8**, 8317-8322 (2014).

122. Lin S., Li X., Wang P., Xu Z., Zhang S., Zhong H., *et al.* Interface designed MoS2/GaAs heterostructure solar cell with sandwich stacked hexagonal boron nitride. *Sci Rep*, **5**,  (2015).

123. Liu X., Galfsky T., Sun Z., Xia F., Lin E.-c., Lee Y.-H., *et al.* Strong light–matter coupling in two-dimensional atomic crystals. *Nat Photon*, **9**, 30-34 (2015).

124. Tan Y., He R., Cheng C., Wang D., Chen Y., Chen F. Polarization-dependent optical absorption of MoS2 for refractive index sensing. *Sci Rep*, **4**, 7523 (2014).

125. Mak K. F., Lee C., Hone J., Shan J., Heinz T. F. Atomically Thin $MoS_2$ : A New Direct-Gap Semiconductor. *Phys Rev Lett*, **105**, 136805 (2010).

126. Splendiani A., Sun L., Zhang Y. B., Li T. S., Kim J., Chim C. Y., *et al.* Emerging Photoluminescence in Monolayer $MoS_2$. *Nano Lett*, **10**, 1271-1275 (2010).

127. Cheng R., Li D., Zhou H., Wang C., Yin A., Jiang S., *et al.* Electroluminescence and photocurrent generation from atomically sharp $WSe_2$/$MoS_2$ heterojunction p–n diodes. *Nano Lett*, **14**, 5590-5597 (2014).

128. Withers F., Del Pozo-Zamudio O., Mishchenko A., Rooney A. P., Gholinia A., Watanabe K., *et al.* Light-emitting diodes by band-structure engineering in van der Waals heterostructures. *Nat Mater*, **14**, 301-306 (2015).

129. Ye Y., Ye Z., Gharghi M., Zhu H., Zhao M., Wang Y., *et al.* Exciton-dominant electroluminescence from a diode of monolayer MoS2. *Appl Phys Lett*, **104**, 193508 (2014).

130. Li D., Cheng R., Zhou H., Wang C., Yin A., Chen Y.*, et al.* Electric-field-induced strong enhancement of electroluminescence in multilayer molybdenum disulfide. *Nat Commun*, **6**, (2015).

131. Lee C.-H., Kim Y.-J., Hong Y. J., Jeon S.-R., Bae S., Hong B. H.*, et al.* Flexible Inorganic Nanostructure Light-Emitting Diodes Fabricated on Graphene Films. *Adv Mater*, **23**, 4614-4619 (2011).

132. Ye Y., Gan L., Dai L., Meng H., Wei F., Dai Y.*, et al.* Multicolor graphene nanoribbon/semiconductor nanowire heterojunction light-emitting diodes. *J Mater Chem*, **21**, 11760-11763 (2011).

133. Han T.-H., Lee Y., Choi M.-R., Woo S.-H., Bae S.-H., Hong B. H.*, et al.* Extremely efficient flexible organic light-emitting diodes with modified graphene anode. *Nat Photon*, **6**, 105-110 (2012).

134. Han N., Viet Cuong T., Han M., Deul Ryu B., Chandramohan S., Bae Park J.*, et al.* Improved heat dissipation in gallium nitride light-emitting diodes with embedded graphene oxide pattern. *Nat Commun*, **4**, 1452 (2013).

135. Amani M., Lien D.-H., Kiriya D., Xiao J., Azcatl A., Noh J.*, et al.* Near-unity photoluminescence quantum yield in MoS2. *Science*, **350**, 1065-1068 (2015).

136. Dou L., Wong A. B., Yu Y., Lai M., Kornienko N., Eaton S. W.*, et al.* Atomically thin two-dimensional organic-inorganic hybrid perovskites. *Science*, **349**, 1518-1521 (2015).

137. Callahan D. M., Munday J. N., Atwater H. A. Solar Cell Light Trapping beyond the Ray Optic Limit. *Nano Lett*, **12**, 214-218 (2012).

138. Ye Y., Wong Z. J., Lu X., Ni X., Zhu H., Chen X.*, et al.* Monolayer excitonic laser. *Nat Photon*, **9**, 733-737 (2015).

139. Wu S., Buckley S., Schaibley J. R., Feng L., Yan J., Mandrus D. G.*, et al.* Monolayer semiconductor nanocavity lasers with ultralow thresholds. *Nature*, **520**, 69-72 (2015).

140. Kang K., Xie S., Huang L., Han Y., Huang P. Y., Mak K. F.*, et al.* High-mobility three-atom-thick semiconducting films with wafer-scale homogeneity. *Nature*, **520**, 656-660 (2015).